\documentclass{aastex61}

\shorttitle{Mid-Infrared Study of M51 AGN at 10--100~pc Scale}
\shortauthors{Ohyama, Matsushita, Oi, and Sun}

\begin{document}

\title{Ground-Based Mid-Infrared Study of the Compton-Thick AGN in M51 at 10--100~pc Scale\footnote{Based in part on data collected at Subaru Telescope, which is operated by the National Astronomical Observatory of Japan.}}

%\author{Youichi Ohyama\altaffilmark{1}, Satoki Matsushita\altaffilmark{1}, Nagisa Oi\altaffilmark{2,3}, and Ai-Lei Sun\altaffilmark{4,1}}
%\affil{$^1$Academia Sinica, Institute of Astronomy and Astrophysics, No.1, Sec. 4, Roosevelt Rd, Taipei 10617, Taiwan, R.O.C.,}
%\affil{$^2$Institute of Space and Astronautical Science, Japan Aerospace Exploration Agency, Yoshinodai 3-1-1, Chuo-ku, Sagamihara, Kanagawa 252-5210, Japan,}
%\affil{$^3$Graduate School of Science and Technology, Kwansei Gakuin University, Gakuen 2-1, Sanda, Hyogo 669-1337, Japan,}
%\affil{$^4$Department of Astrophysics, Princeton University, Princeton, NJ 08540}

\author{Youichi Ohyama}
\affiliation{Academia Sinica, Institute of Astronomy and Astrophysics, No.1, Sec. 4, Roosevelt Rd, Taipei 10617, Taiwan, R.O.C.}

\author{Satoki Matsushita}
\affiliation{Academia Sinica, Institute of Astronomy and Astrophysics, No.1, Sec. 4, Roosevelt Rd, Taipei 10617, Taiwan, R.O.C.}

\author{Nagisa Oi}
\affiliation{Institute of Space and Astronautical Science, Japan Aerospace Exploration Agency, Yoshinodai 3-1-1, Chuo-ku, Sagamihara, Kanagawa 252-5210, Japan}
\affiliation{Graduate School of Science and Technology, Kwansei Gakuin University, Gakuen 2-1, Sanda, Hyogo 669-1337, Japan}

\author{Ai-Lei Sun}
\affiliation{Academia Sinica, Institute of Astronomy and Astrophysics, No.1, Sec. 4, Roosevelt Rd, Taipei 10617, Taiwan, R.O.C.}
\affiliation{Department of Astrophysics, Princeton University, Princeton, NJ 08540}

\begin{abstract}
We performed near-diffraction-limited ($\simeq 0\farcs 4$ FWHM) $N$-band imaging of one of the nearest Active Galactic Nucleus (AGN) in M51 with 8.2m Subaru telescope to study the nuclear structure and spectral energy distribution (SED) at 8--13~$\mu$m.
We found that the nucleus is composed of an unresolved core (at $\simeq 13$~pc resolution, or intrinsic size corrected for the instrumental effect of $<6$~pc) and an extended halo (at a few tens~pc scale), and each of their SEDs is almost flat.
We examined the SED by comparing with the archival {\it Spitzer} IRS spectrum processed to mimic our chopping observation of the nucleus, and the published radiative-transfer model SEDs of the AGN clumpy dusty torus.
The halo SED is likely due to circumnuclear star formation showing little Polycyclic Aromatic Hydrocarbon (PAH) emission due to the AGN.
The core SED is likely dominated by the AGN because of the following two reasons.
Firstly, the clumpy torus model SEDs can reproduce the red mid-infrared continuum with apparently moderate silicate 9.7~$\mu$m absorption.
Secondly, the core 12~$\mu$m luminosity and the absorption-corrected X-ray luminosity at 2--10~keV in the literature follow the mid-infrared--X-ray luminosity correlation known for the nearby AGNs including the Compton-thick ones.
\end{abstract}

\keywords{galaxies: active --- galaxies: individual (M51) --- galaxies: nuclei --- infrared: galaxies}

\section{INTRODUCTION}\label{intro}

The dusty torus around Active Galactic Nucleus (AGN) is a key component in a unification theory of AGNs (\citealt{antonucci93}; see also \citealt{netzer15} for a recent review).
Such torus is originally introduced to hide broad line region, another key component in the unification theory, from observers' line-of-sight for type-2 AGNs in the optical and near-infrared (NIR) wavelengths.
In the unification theory, the torus is also responsible for major characteristics of AGNs at X-ray and mid-infrared (MIR) wavelengths.
X-ray emission, especially in the soft X-ray band, from the central part of the AGN is absorbed by gas within the torus.
The observed absorption is larger for type-2 AGNs due to the dusty torus intercepting along the line-of-sight, and most X-ray emission below 10~keV is absorbed in the extreme Compton-thick case ($N_{\rm H}>1.25\times 10^{24}$~cm$^{-2}$, where $N_{\rm H}$ is the hydrogen column density).
MIR emission from the AGN is dominated by thermal emission of the dusty torus heated by the AGN.
Silicate feature around 9.7~$\mu$m is often seen either in absorption or emission when the torus is seen edge-on or pole-on, respectively, by the observers.

Recent MIR studies have provided us with much more realistic view of the central part of the AGNs.
{\it Spitzer} studies of nearby Compton-thick AGNs have shown that even Compton-thick AGNs, especially low-luminosity ones, often show only modest--moderate silicate absorption at 9.7$~\mu$m (e.g., \citealt{hao07,goulding12}).
Classical smooth torus model, such as one of \cite{pier92}, predicts deeper absorption in proportion to the X-ray absorption column density.
On the other hand, if the torus is made of collection of clouds, each cloud is heated to $\sim 300$~K to emit MIR emission while absorbing the background light when the foreground cloud is cooler than the one behind.
The radiation transfer effect among the clouds significantly reduces the net silicate absorption even when the torus is seen edge-on \citep{nenkova02,honig06,nenkova08a,nenkova08b,honig10,stalevski12,stalevski16}.
Meanwhile, recent MIR interferometric studies of nearby AGNs have started to directly reveal the dust distribution in the vicinity of the AGNs at parsec scales.
In some best studied AGNs, extended optically-thin dust emission elongated toward the system's polar direction (e.g., direction of the extended narrow line region or outflow) is typically found in addition to the compact disk-like component (e.g., \citealt{raban09,honig12,tristram12,honig13,tristram14,lg16}; see also \citealt{asmus16} for the single dish study; see \citealt{netzer15} for a review).
Such extended polar emission is clearly inconsistent with the classical idea of the dusty torus in the unification theory, and its nature is under debate.
Some proposed ideas are that it originates from the inner funnel of an extended dust distribution above and below the torus and/or the dusty outflow within the ionizing cone that is radiatively driven from the inner wall of the compact dusty torus (e.g., \citealt{honig12,keating12,roth12,honig13,tristram14}).

M51 (NGC~5194) is a very nearby (7.1~Mpc; \citealt{tv06}; $1''$ corresponds to 34.4~pc) Compton-thick AGN.
It is one of the nearest AGNs, and is even closer than the best studied Compton-thick AGN NGC~1068 (14.4~Mpc).
Although the AGN in M51 (0.07~Jy for the unresolved core component at a resolution of 0\farcs 4 full-width at half maximum (FWHM) at 11.2~$\mu$m; \S \ref{radial_profile_analysis}) is much fainter than that of NGC~1068 (8.7~Jy within a central 0\farcs 4 aperture at 11.6~$\mu$m; \citealt{mason06}), it has been studied in great detail, thanks to its proximity.
The hard X-ray emission from the AGN was discovered by {\it GINGA} \citep{makishima90}, and its Compton-thick spectra have been analyzed in more detail by using the succeeding X-ray satellites \citep{terashima98,fukazawa01,panessa06,brightman11,lamassa11}.
Extended AGN-related activities such as radio jet (e.g., \citealt{crane92,bradley04,rampadarath15}), optical narrow line region nebula (e.g., \citealt{cecil84,bradley04}), and soft X-ray nebula \citep{terashima01} have been known.
Torus-like structures at $\sim 100$~pc scale have been claimed in the 1990s by several authors, although they are not likely the torus in light of recent advanced high-resolution infrared interferometric studies of nearby AGNs (e.g., \citealt{kishimoto11,burtscher13}).
Those include disk-like rotating dense molecular gas cloud at $\sim 100$~pc scale with HCN ($J=1-0$) emission \citep{kohno96} and ``X''-shaped nuclear dust lanes at several tens~pc scale in {\it HST} optical images \citep{grillmair97}.
Recently, \cite{satoki07} found that most molecular gas is associated with circumnuclear outflowing structures at several tens~pc scale, with little nuclear concentration.
They argued that the hydrogen column density toward the nucleus is apparently much smaller (by more than two orders of magnitude) than the expectation from the Compton-thick X-ray properties (see also \citealt{satoki15}; \S \ref{discussion_torus}).

High spatial-resolution MIR imaging of one of the nearest AGNs in M51 would enable us to examine details of the AGN structure and its circumnuclear region.
In the $N$-band (8--13~$\mu$m), ground-based 8--10~m telescopes can achieve diffraction-limited resolution of 0\farcs 3--0\farcs 4, or $\sim 10$~pc at a distance of M51.
Note that the most powerful high-resolution imaging/spectral mapping machine at MIR, Very Large Telescope Interferometer (VLTI) with MIDI (MID-infrared Interferometric) instrument, cannot observe this galaxy due to its declination.
In addition, space-based MIR imaging and spectroscopy provide the best sensitivity over the larger field of view and larger wavelength coverage at modest spatial resolution, enabling detailed spectral analysis of the AGN and its circumnuclear region.
Therefore, we performed Subaru MIR imaging observation and analyzed the archival {\it Spitzer} data of the AGN in M51 to characterize the AGN inner structure and effect of the AGN on the circumnuclear region.

\section{COMICS DATA AND ANALYSIS}

\subsection{Observation and Image Processing}\label{observations}

\begin{figure}
\begin{center}
\includegraphics[scale=0.7,angle=-90]{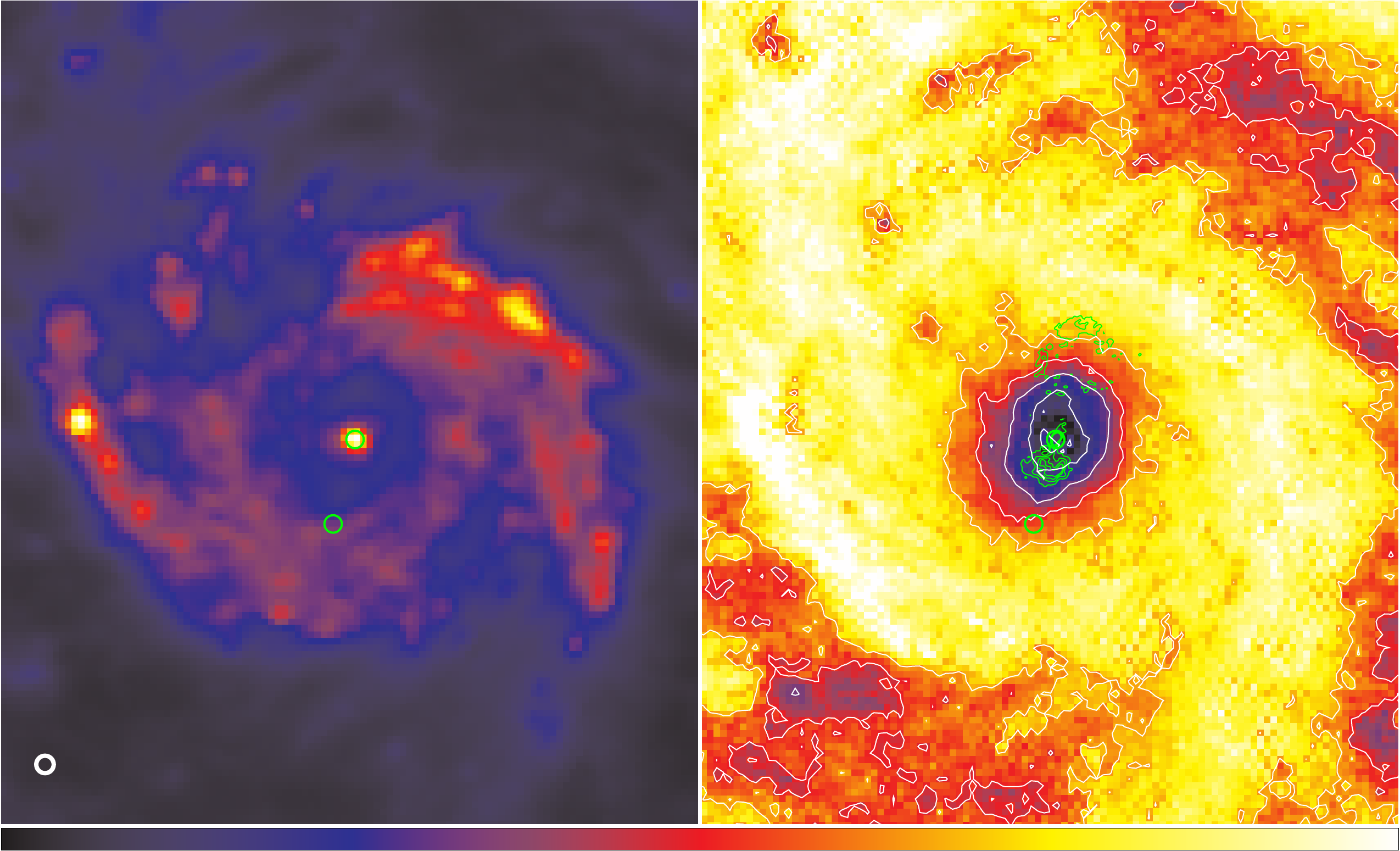}
\caption{
IRAC CH4 (8.0~$\mu$m; left) and CH4/CH3 flux ratio (8.0~$\mu$m/5.6~$\mu$m; right) images of the central region of M51 in linear scale.
Dark blue color in the right panel indicates smaller ratio as indicated by a color bar, and the contours (white) are drawn at the flux ratios of 1.5, 1.8, 2.1, and 2.4.
The representative flux ratios are $\simeq 1.5$ near the nucleus, $\simeq 2.7$ along the nuclear ring and inner arms, and $\simeq 2.0$ in the inter-arm region.
The IRAC resolution ($\simeq 2\farcs 0$) is shown with a white circle in the left panel.
Our COMICS beam positions (nucleus and ``sky'' at 10\arcsec~off toward PA$=166^{\rm \circ}$) are marked with green circles in both panels.
A 6~cm VLA radio map of \cite{crane92} is overlaid with green contours in the right panel.
\label{findingchart}}
\end{center}
\end{figure}

Imaging observation in the $N$-band was made with COMICS \citep{comics} at the Subaru telescope on a night of April 20, 2011 (UTC).
We employed four filters, two medium-band continuum filters: ``N8.8'' ($\lambda_{\rm cen}=8.8$~$\mu$m; $\Delta \lambda=0.8$~$\mu$m) and ``N10.5'' ($\lambda_{\rm cen}=10.5$~$\mu$m; $\Delta \lambda=1.0$~$\mu$m), and two narrow-band filters for spectral features: ``UIR11.2'' ($\lambda_{\rm cen}=11.24$~$\mu$m; $\Delta \lambda=0.60$~$\mu$m) and ``[Ne~{\sc ii}]'' ($\lambda_{\rm cen}=12.81$~$\mu$m; $\Delta \lambda=0.21$~$\mu$m), covering most of the entire $N$-band window altogether.
One pixel corresponds to 0\farcs 13 on the sky.
The observation was made in a standard procedure for ground-based MIR observations with secondary mirror chopping.
We chopped between the nucleus and ``sky'' region at a frequency of about 0.5~Hz, while keeping the M51 nucleus on-chip at both positions (Figure~\ref{findingchart}).
The sky is 10\arcsec~off the nucleus toward PA$=166^{\rm \circ}$, and is not a true background sky.
It is rather located at relatively smooth region within M51 between the nucleus and the most inner star-forming ring.
While the secondary mirror stays at one chop position for over $\simeq 1$ second, the detector array is read out multiple times, and the instrument computer stacks them over that period to create a frame, which is a unit for us to process off-line.
When the secondary mirror was chopped toward the nucleus position, the telescope was guided with an independent offset guiding system by using a nearby bright star.
We employed the same chopping pattern both for M51 and photometric standard stars, and we measured the chop distance by using the standard star observations.
We did not employ frequent telescope nodding, because this chopping method alone gives relatively flat ``background'' region when subtracting frames between the two chopping positions.
Note that we detected only a compact nucleus at a scale of about one arcsec with all filters, and everywhere else can be considered effectively as background (\S \ref{radial_profile_analysis}).
However, we nodded the telescope at much longer timescale of $\sim 30$ minutes.
This is to reduce effect of bad pixels and pixel-dependent calibration errors in the final stacked images.
We measured the nod distance by using partly stacked images of the M51 nucleus itself taken at the same telescope nod positions.

Flux calibration and monitoring of the photometric/seeing conditions were made based on photometric standard star observations.
We choose HD107274 from \cite{cohen99} since it is closest to M51 ($\simeq 12^{\rm \circ}$ away).
We also observed HD120993, the second closest star in the list, when HD107274 went below the telescope elevation of $30^{\rm \circ}$.
Their in-band fluxes are summarized in Table~\ref{obslog}.
The standard stars were observed immediately before or after (or both) each observing session of about 1--2 hours, during which M51 had been observed continuously with one filter, by using the same filter and the detector readout pattern for the M51 observation.
We measured variation of standard star count rates during the night, and found no significant change ($\simeq 4$\% in the [Ne~{\sc ii}] image and less in the other band images) during the night.
We also measured seeing variation during the night by using the standard star observations, and found no significant change of the FWHM size of the stars.
Table~\ref{obslog} summarizes the observations as well as the photometric and seeing conditions during the night.

\begin{table}[ht]
\caption{Observation logs and observing conditions.\label{obslog}}
\begin{tabular}{cccccccccccc}
\tableline\tableline
\multicolumn{2}{c}{Filter} & Diff.-limited\tablenotemark{a} & Exposure & \multicolumn{2}{c}{Standard Star} & $N_{\rm obs}^{\rm std}$\tablenotemark{b} & \multicolumn{4}{c}{Seeing FWHM\tablenotemark{c}} & Photometry\tablenotemark{d} \\
Name & $\lambda_{\rm cen}$/$\Delta \lambda$ & FWHM & & HD number & Flux & & mean & error & min & max & variation\\
 & ($\mu$m) & ($''$) & (s) & & (Jy) & & ($''$) & ($''$) & ($''$) & ($''$) & (\%)\\
\tableline
N8.8                   & 8.8/0.8       & 0.27 & 401   & 107274             & 18.3         & 1 & 0.38 & 0.004 & \nodata  & \nodata     & \nodata  \\
N10.5                 & 10.5/1.0     & 0.32 & 2406 & 107274             & 14.5         & 2 & 0.32 & 0.01 & 0.31 & 0.34 & 0.1 \\
UIR11.2              & 11.24/0.60 & 0.35 & 4390 & 107274/120993 & 12.7/32.7 & 4 & 0.37 & 0.01 & 0.34 & 0.40 & 3    \\
{\rm [Ne~{\sc ii}]} & 12.81/0.21 & 0.39 & 4122 & 107274             & 10.0         & 4 & 0.43 & 0.02 & 0.41 & 0.45 & 4    \\
\tableline
\end{tabular}
\tablenotetext{a}{Theoretical diffraction-limited resolution.}
\tablenotetext{b}{Number of standard star observations during the night.}
\tablenotetext{c}{Mean and its error, minimum, and maximum seeing size during the night measured with the standard stars.}
\tablenotetext{d}{Count variation in percent of the same standard star observed multiple times during the night.}
\end{table}

We performed a standard image processing for ground-based MIR imaging observations with secondary mirror chopping by using our own IDL program.
This includes subtraction of chop-paired frames, flat fielding, image shifting for the secondary-mirror chopping and telescope nodding, and stacking.
We also checked photometric conditions of all individual frames, by measuring the total flux (before subtracting the chop-paired frames) and pixel-to-pixel flux variation (after subtracting the chop-paired frames and flat-fielding) in the background region away from the M51 nucleus.
Both quantities rapidly increase when cloud passes the telescope beam.
We removed a small fraction ($\sim 5$\% for the UIR11.2 image and less for other filter images) of frames taken under poor sky conditions before stacking.
We finally detected a compact structure at the M51 nucleus at $\sim 1''$ scale in all four bands (see below for the radial profile).
No other notable structures such as elongated disks or extended arms are detected.

\subsection{PSF Characterization and Radial Profile Analysis of M51 Nucleus}\label{radial_profile_analysis}

\begin{figure}
\plotone{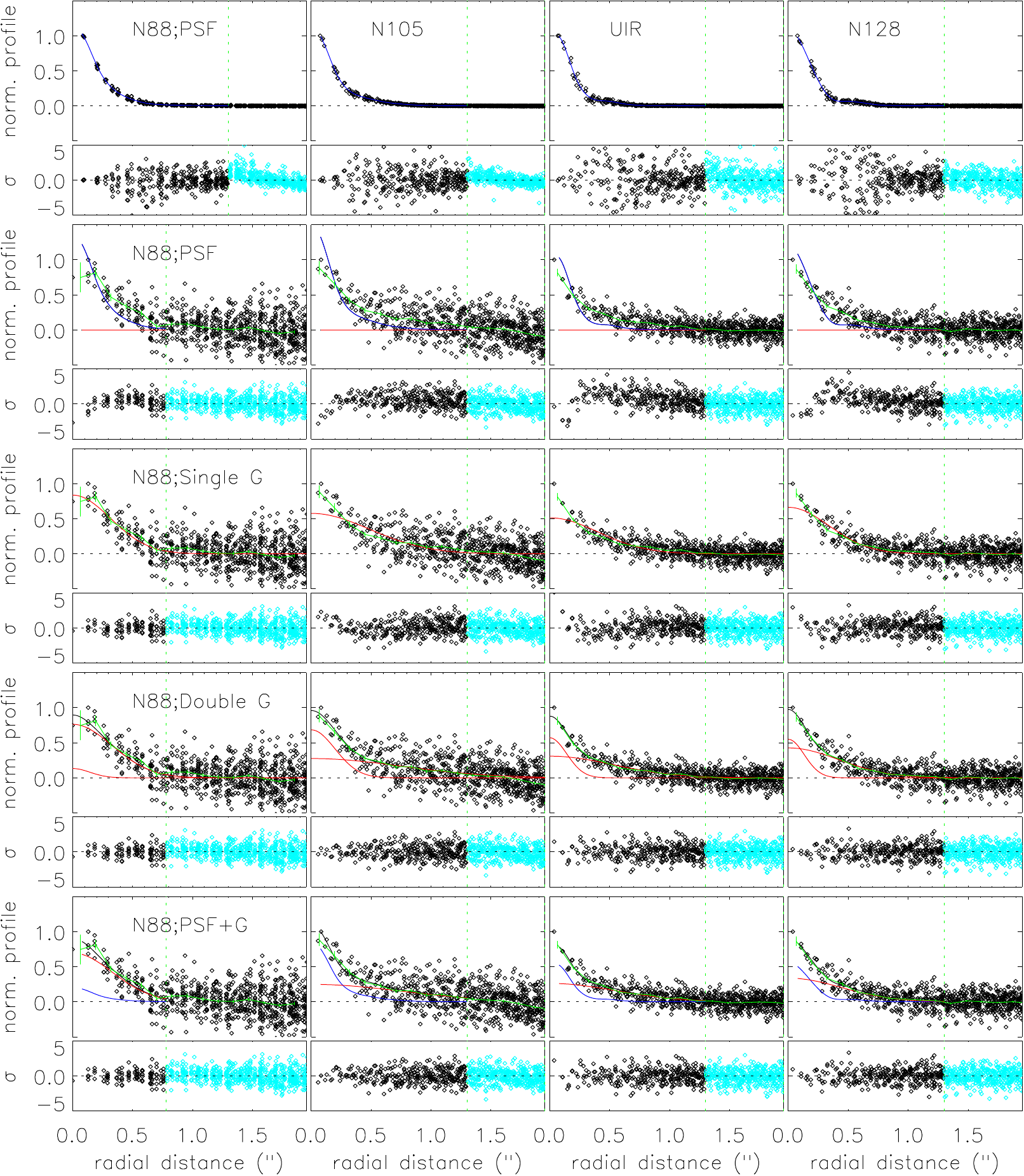}
\caption{
Results of the radial profile fitting of the M51 nucleus.
The radial profiles of the standard star and M51 nucleus are shown in top and the following four rows, respectively.
The radial profiles are normalized to be unity at the peak, and are shown as a function of radius in units of arcsec.
The radial profiles of N8.8, N10.5, UIR11.2, and [Ne~{\sc ii}] filter images are shown in the left most, second, third, and the right most columns, respectively.
For the M51 nucleus, both pixel-based observed profile (black diamonds) and binned profiles within individual radial bins with error bars for one-sigma data scattering (green) are shown.
We used the pixel-based profiles and the binned profiles for the fitting and visual guide, respectively.
In each panel, a vertical green dashed line indicates boundary between the outer region for background subtraction and the object aperture for profile fitting and photometry.
Fitting results with the models of PSF-only, single Gaussian, double Gaussian, and PSF+Gaussian components are shown in the second, third, fourth, and bottom rows, respectively.
The best-fit models are overlaid in the upper sub-panel, and the noise-normalized residual (sigma=(observation $-$ best-fit model)/noise) profile within the aperture (black) and the noise-normalized observation data within the background region (cyan) are shown in the lower sub-panel.
Here, the noise is one-sigma uncertainty of the observed (and normalized) profile data.
For models including PSF (top, second, and bottom rows), the best-fit PSF component is shown in blue.
For the dual component models (fourth and fifth rows), the best-fit Gaussian components are shown in red, and sums of the two components (either PSF+Gaussian or double Gaussian) are shown in black lines.
\label{radialfit_results}}
\end{figure}

The Point Spread Function (PSF) is carefully characterized with standard star images.
Because the PSF was stable during the night (\S \ref{observations}), we stacked standard star images for each filter to derive the PSF radial profile (top row of Figure~\ref{radialfit_results}).
Note that, in order to avoid difficulties in handling non-axisymmetric PSFs\footnote{
COMICS typically shows the non-axisymmetric PSF, most notably the slight trefoil-like PSF core due to non-axisymmetric telescope structure such as the prime mirror supporting structure.
Such PSF rotates on the focal plane as the telescope on the altazimuth mount tracks the object in the sky and, therefore, simply stacked images show their own complicated averaged PSFs.}
, our analysis of the source structure and photometries are entirely based on the azimuthally-averaged radial brightness profiles (hereafter, the radial profiles) of the standard star and M51 images after simply stacking the frames without considering the PSF rotation.
We found that N10.5, UIR11.2, and [Ne~{\sc ii}] images are almost diffraction-limited, while the resolution of the N8.8 image is worse (Table~\ref{obslog}).
For the N8.8 and N10.5 images, we smoothed both standard star and M51 images in the same way to improve the signal-to-noise ratio by convolving a Gaussian kernel of 1.5 pixel (or 0\farcs 19) FWHM.
We use these smoothed images for the rest of this paper.
The PSF radial profiles are apparently made of a sharp core and an extended structure around it, and they are modeled with B-Spline function.

We performed detailed radial profile analysis of the M51 nucleus.
We fit the M51 radial profile with the following models: single PSF (for an unresolved core only), single Gaussian, double Gaussian (for a compact core + extended halo), and PSF+Gaussian (for an unresolved core + extended halo).
Goodness of the fits is evaluated by using reduced $\chi^2$ and inspection of fit residual (observation $-$ best-fit model) patterns.
Note that the reduced $\chi^2$ is not very sensitive to modeling error near the core because there are much larger number of pixels to fit at outer radius.
The fit results are shown in Figure~\ref{radialfit_results} and Table~\ref{table_radialfit_results}.

\begin{table}[ht]
\caption{Results of radial profile fitting and photometry of the M51 nucleus.\label{table_radialfit_results}}
\begin{tabular}{ccccccccccc}
\tableline\tableline
Filter & PSF & \multicolumn{4}{c}{reduced $\chi^2$ of each model} & \multicolumn{3}{c}{Best fit source size FWHM} & \multicolumn{2}{c}{Photometry\tablenotemark{e}} \\
 & FWHM\tablenotemark{a} & PSF & G\tablenotemark{b} & G+G\tablenotemark{b} & PSF+G\tablenotemark{b} & G\tablenotemark{b} & core\tablenotemark{c} of G+G & halo\tablenotemark{d} of PSF+G & core+halo & core \\
 & ($''$) & & & & & ($''$) & ($''$) & ($''$) & (mJy) & (mJy) \\
\tableline
N8.8                    & $0.45 \pm 0.01$\tablenotemark{f} & 0.98 & 0.64 & 0.65 & 0.64 &  $0.74\pm 0.05$\tablenotemark{f} & $0.33\pm 0.64$\tablenotemark{f} & $0.77\pm 0.10$\tablenotemark{f} & $15.57\pm 1.30$ & $1.66\pm 2.96$ \\
N10.5                 & $0.40 \pm 0.00$\tablenotemark{f} & 1.37 & 1.11 & 1.00 & 1.01 &$1.14\pm 0.06$\tablenotemark{f} & $0.46\pm 0.07$\tablenotemark{f} & $1.77\pm 0.24$\tablenotemark{f} &$23.05\pm 1.89$ & $5.71\pm 0.85$ \\
UIR11.2              & $0.39 \pm 0.02$                            & 1.85 & 1.21 & 1.03 & 1.04 & $0.94\pm 0.03$ & $0.33\pm 0.04$ & $1.28\pm 0.08$ & $35.44\pm 2.33$ & $7.21\pm 0.86$ \\
{\rm [Ne~{\sc ii}]} & $0.42 \pm 0.01$                             & 1.43 & 1.06 & 0.98 & 0.98 & $0.79\pm 0.03$ & $0.31\pm 0.05$ & $1.03\pm 0.08$ & $65.89\pm 5.55$ & $15.47\pm 2.62$ \\
\tableline
\end{tabular}
\tablenotetext{a}{Measured FWHM size of the PSF radial profile used for the profile fitting of the M51 nucleus.
For N8.8 and N10.5, both stacked standard star and M51 images are smoothed in the same ways to improve the signal-to-noise ratio, and the profile fitting was made on the radial profiles of those smoothed images. The PSF sizes here for N8.8 and N10.5 are the ones after the smoothing. See text for the smoothing.}
\tablenotetext{b}{G = single Gaussian component.}
\tablenotetext{c}{The Gaussian component for the core of the double Gaussian model.}
\tablenotetext{d}{The Gaussian component for the extended halo component of the PSF+Gaussian model.}
\tablenotetext{e}{Based on the PSF+Gaussian model.}
\tablenotetext{f}{After applying the smoothing}
\end{table}

First, we fit the M51 nucleus radial profile with the single component models.
In the N10.5, UIR11.2, and [Ne~{\sc ii}] images, we found that a fit with a single PSF clearly fails with large reduced $\chi^2$ ($>1.0$) and the systematic residual pattern of negative and positive deviation from zero near the nucleus ($r\lesssim$ FWHM(PSF)) and outward ($r\gtrsim$ FWHM(PSF)), respectively.
This indicates presence of additional extended component.
The similar, but less pronounced, systematic deviation of the residual pattern is also seen in the N8.8 image, although the reduced $\chi^2$ is smaller in this filter image.
For the N10.5, UIR11.2, and [Ne~{\sc ii}] images, a fit with a single Gaussian is much better with smaller reduced $\chi^2$ ($\gtrsim 1.0$), but the fit result shows systematic residual pattern of positive and negative deviation from zero near the nucleus ($r\lesssim$ FWHM(PSF)) and outward ($r\gtrsim$ FWHM(PSF)), respectively.
This indicates presence of additional compact core.
Again, the similar, but less pronounced, systematic deviation of the residual pattern is also seen in the N8.8 image, although the reduced $\chi^2$ is smaller in this filter image.
We found that the best fit single Gaussian profile is much wider than the PSF in all filters, i.e., the M51 nucleus is resolved in all filters.
Note that the FWHMs of the single Gaussian model are consistent with previous measurements by \cite{asmus14} under poor sky conditions with Michelle instrument at Gemini North telescope at the similar wavelengths.

We then fit the M51 nucleus radial profile with double (core+halo) component models.
In the N10.5, UIR11.2, and [Ne~{\sc ii}] images, we found that both double Gaussian and PSF+Gaussian models are much better than the single component models in terms of both smaller reduced $\chi^2$ ($\simeq 1$) and little fit residual patterns.
In the double Gaussian model, widths of the compact Gaussian component (0\farcs 3--0\farcs 4 FWHM) are roughly comparable to those of the PSFs, supporting an idea of presence of an unresolved core at the nucleus.
Such unresolved core is naturally expected for this AGN.
Therefore, we prefer a model of PSF+Gaussian for the three filter images.
To estimate the upper limit of the intrinsic core size corrected for the instrumental effect, we assumed that the core intrinsically shows a Gaussian profile and the observed core is actually a Gaussian-smoothed PSF.
We fit the UIR11.2 and [Ne~{\sc ii}] images with the Gaussian-smoothed PSF for the core and the best-fit Gaussian component for the extended component, and measured $\chi^2$ values for various Gaussian core sizes.
We found that the one sigma upper limit of the core size (FWHM) is 8~pc and 6~pc in UIR11.2 and [Ne~{\sc ii}] images, respectively.
We adopt 6~pc as the best estimate of the intrinsic core size upper limit.
As for the N8.8 image, the single Gaussian model is as good as the double component models, and the single Gaussian fit indicates that this component is extended.
We thereby suspect that the N8.8 image is also made of an unresolved core+halo structure as in the case of other filter images, although we cannot confirm presence of the core component in this band.

\begin{figure}
\plotone{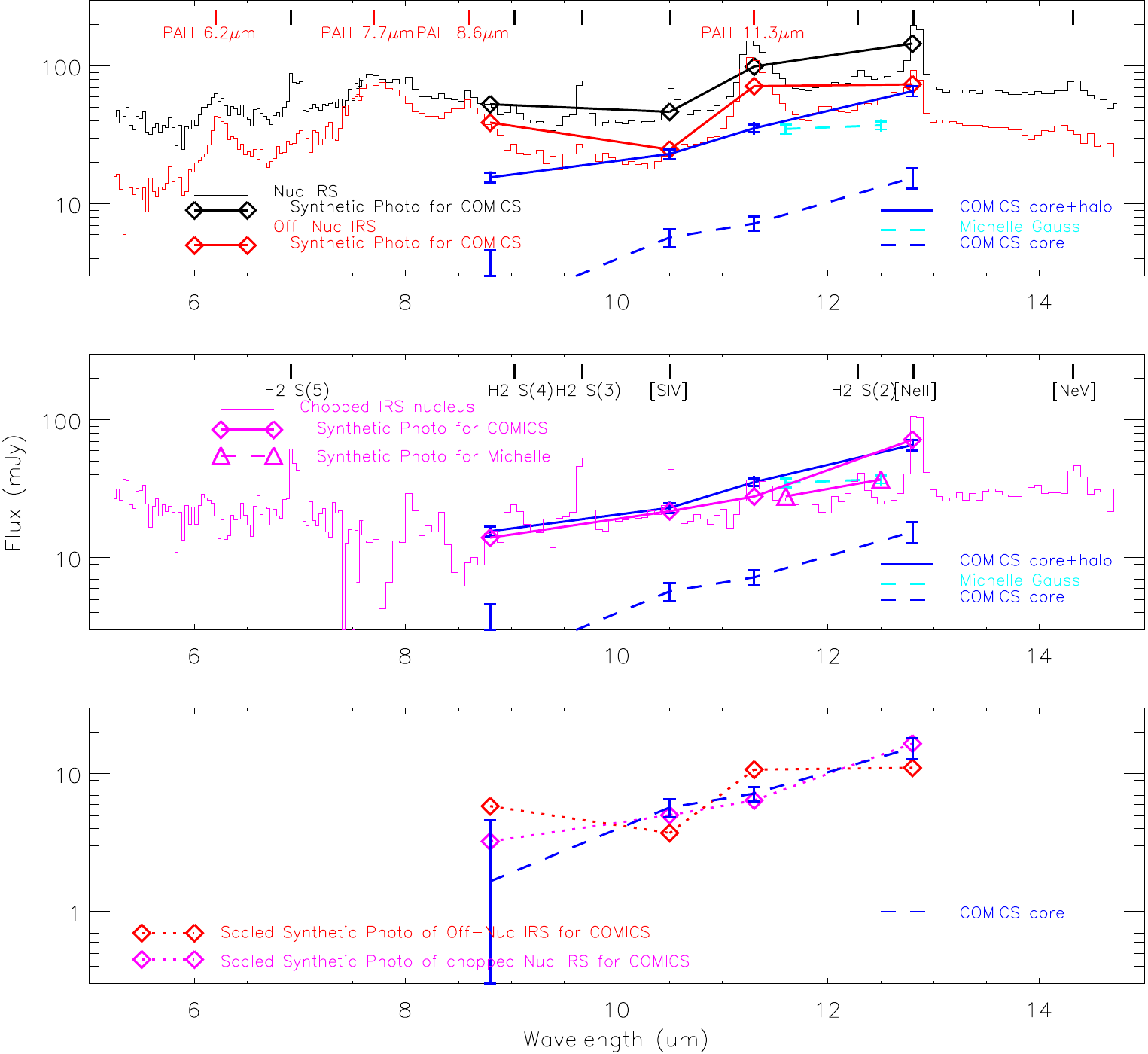}
\caption{
The nucleus and off-nucleus (spectro)photometries of M51.
(top)
The nuclear (thin black line) and off-nuclear (thin red line) IRS spectra and their corresponding synthetic photometries for the COMICS filters (diamonds) are shown.
Wavelengths of the detected emission lines and the PAH features are marked, and the PAH features are identified.
The observed COMICS photometries of the core+halo (blue solid) and the core (blue broken) components are overlaid.
The observed Michelle photometries from \cite{asmus14} (with Gaussian profile fitting; cyan) are compared.
(middle)
The chopped nuclear IRS spectrum, which is the difference between the two IRS spectra in the top panel, is shown in magenta, and the corresponding synthetic photometries for the COMICS and Michelle filters are overlaid in magenta diamonds and triangles, respectively.
Wavelengths of the detected emission lines are identified.
The same observed COMICS and Michelle photometries in the upper panel are also shown with the same colors and symbols for comparison.
(bottom)
The COMICS core photometries (blue broken) are compared with the synthetic photometries of the scaled IRS off-nuclear spectrum (red diamonds) and of the scaled chopped nuclear IRS spectrum (magenta diamonds).
Here, the scaling on the chopped nuclear IRS spectrum is based on the typical flux ratio between the core and halo COMICS photometries (0.3; \S \ref{radial_profile_analysis}), and the scaling on the off-nuclear IRS spectrum is arbitrary to match the original synthetic photometries of the off-nuclear IRS spectrum to the observed COMICS core photometries.
\label{sed}}
\end{figure}

We use the profile fitting results with the PSF+Gaussian model to do photometry of each core and halo component of the M51 nucleus in all four filter images (Table~\ref{table_radialfit_results}).
Figure~\ref{sed} shows the resultant SEDs of the M51 nucleus.
The core+halo SED is rather flat with slightly increasing flux at longer wavelength.
Flux ratios between the core and halo are about 0.3 in all filters, implying that each of the core and halo SEDs shows the similar flat SED.

\subsection{Comparison with Michelle Photometries}\label{comp_michelle}

We compare the COMICS core+halo photometries with the previous measurements by \cite{asmus14}.
They used Michelle instrument at the Gemini North telescope, and employed two medium-band ``silicate'' filters: ``Si-5'' ($\lambda_{\rm cen}=11.6$~$\mu$m; $\Delta \lambda=1.1$~$\mu$m) and ``Si-6'' ($\lambda_{\rm cen}=12.5$~$\mu$m; $\Delta \lambda=1.2$~$\mu$m).
Note that the Si-6 filter includes [Ne~{\sc ii}] but its width (1.2~$\mu$m) is much wider than the [Ne~{\sc ii}] filter of COMICS (0.21~$\mu$m).
Because of poor seeing, they simply reported that the nucleus could be marginally resolved.
We compare our core+halo photometries with their single Gaussian fitting photometries (top and middle panels of Figure~\ref{sed}).
This is because they claimed that their PSF-fitting photometry is not reliable under poor seeing conditions but the Gaussian fitting photometry is reliable to measure the total flux from the nuclear region.
We found that the COMICS UIR11.2 (11.2~$\mu$m) flux is consistent with the Michelle Si-5 (11.6~$\mu$m) flux, whereas the COMICS [Ne~{\sc ii}] (12.8~$\mu$m) flux is notably higher than the Michelle Si-6 (12.5~$\mu$m) flux.
We examine this difference in combination with the {\it Spitzer} data in \S \ref{results_core_halo} below.

\section{{\it SPITZER} DATA AND ANALYSIS}\label{spitzer_data_analysis}

M51 was observed by {\it Spitzer} as a part of the SINGS survey \citep{sings}, and a number of papers were already published for this galaxy (e.g., \citealt{brunner08,mm09a,mm09b,ds10,schinnerer13}).
We retrieved both CH3 (5.8~$\mu$m) and CH4 (8.0~$\mu$m) IRAC \citep{irac} mosaic images, and SL1 and SL2 (covering $5-15$~$\mu$m in total) IRS \citep{irs} spectral cubes (reconstructed from slit-scanning data) from the SINGS DR2 data distribution.
We aligned both SL1 and SL2 cubes to form a single cube for our study.
Both CH3 and CH4 images trace hot dust continuum and Polycyclic Aromatic Hydrocarbon (PAH) emission features (e.g., \citealt{schinnerer13}), and a compact (but slightly resolved) nucleus and surrounding ring/spiral-arm structures are evidently seen at a few tens arcsec scale in these bands (Figure~\ref{findingchart}).
They show a clear contrast to both IRAC CH1 (3.6~$\mu$m) and CH2 (4.5~$\mu$m) images, which trace the stellar distribution showing an extended bulge structure around the nucleus \citep{schinnerer13}.
The IRS cubes also show similar structures as seen in the IRAC images, but the nucleus is not spatially resolved with IRS (see \citealt{brunner08} for full details).

To enable direct comparison between the IRS spectra and the COMICS photometries, we extracted IRS spectra from the cubes both on the nucleus and at the off-nucleus ``sky'' region of the COMICS chopping observation (\S \ref{observations}; top panel of Figure~\ref{sed}).
Aperture size of each region is $3 \times 3$ pixels of the IRS cube ($5\farcs 6 \times 5\farcs 6$).
This size is large enough to include most of the flux from the unresolved IRS nucleus, but it is much larger than typical scale of the nucleus seen with COMICS (\S \ref{radial_profile_analysis}).
These spectra are very similar to those presented by \cite{ds10} (their Figure 3), although their spectra are extracted with slightly different extraction box size ($3\farcs 6 \times 2\farcs 7$) and their off-nuclear spectrum is extracted at the different location.
A ``chopped'' nuclear IRS spectrum was also made by subtracting the ``sky'' spectrum from the nuclear one (middle panel of Figure~\ref{sed}).
Synthetic photometries on these three IRS spectra for the COMICS filters are made to compare with the COMICS photometries.

The nuclear and off-nuclear IRS spectra show numbers of differences (top panel of Figure~\ref{sed}).
The off-nuclear spectrum is similar to that of typical star-forming region (e.g., \citealt{smith07}; see also \citealt{ds10}).
On the other hand, as originally proven by \cite{ds10}, the nuclear spectrum shows stronger [Ne~{\sc ii}]$\lambda 12.81$~$\mu$m, [Ne~{\sc v}]$\lambda 14.32$~$\mu$m, [S~{\sc iv}]$\lambda 10.51$~$\mu$m, and pure-rotational H$_2$ lines ($0-0$ $S$(5) at 6.91~$\mu$m, $0-0$ $S$(4) at 9.03~$\mu$m, $0-0$ $S$(3) at 9.67~$\mu$m, and $0-0$ $S$(2) at 12.28~$\mu$m), with weaker PAH 6.2~$\mu$m and 7.7~$\mu$m features (see also \citealt{dudik07,ps10}).
The chopped nuclear IRS spectrum is quite different from both nuclear and off-nuclear spectra (middle panel of Figure~\ref{sed}).
It is flat-continuum dominated with little PAH features, and shows only fine structure lines such as [Ne~{\sc ii}], [Ne~{\sc v}], and pure-rotational H$_{2}$ lines.
We found that the [Ne~{\sc ii}]/[Ne~{\sc v}] flux ratio is about 4.

The IRAC CH4/CH3 flux ratio image shows a trend within a few tens arcsec from the nucleus, in a sense that the ratio is smallest at the nucleus and it gradually becomes larger with increasing radius from the nucleus (Figure~\ref{findingchart}).
Beyond this radius, the ratio image shows structures such as a circumnuclear ring and inner arms, where the ratio is relatively enhanced.
Because the nuclear IRS spectrum shows weaker PAH 7.7~$\mu$m feature (top panel of Figure~\ref{sed}), which is usually prominent in the CH4 band in star-forming regions, the smaller CH4/CH3 flux ratio around the nucleus is likely due to weaker PAH features.

\section{RESULTS}\label{results}

\subsection{Core+Halo SED}\label{results_core_halo}

The COMICS core+halo SED of the M51 nucleus is quite different from typical SED of star-forming objects and the IRS off-nuclear spectrum of M51 (top panel of Figure~\ref{sed}), and is instead consistent with the chopped nuclear IRS spectrum showing little PAH features (middle panel of Figure~\ref{sed}).
When compared to the Michelle photometries of the M51 nucleus \citep{asmus14}, the COMICS UIR11.2 (11.2~$\mu$m) flux is consistent with the Michelle Si-5 (11.6~$\mu$m) flux, whereas the COMICS [Ne~{\sc ii}] (12.8~$\mu$m) flux is notably higher than the Michelle Si-6 (12.5~$\mu$m) flux (\S \ref{comp_michelle}).
In an attempt to explain the difference, we made the synthetic photometries of the chopped nuclear IRS spectrum for the two Michelle filters (top and middle panels of Figure~\ref{sed}).
We found that the COMICS [Ne~{\sc ii}] flux is larger than the Si-6 flux because the [Ne~{\sc ii}] filter just covers the [Ne~{\sc ii}] line whereas the Si-6 filter also covers the adjacent continuum due to its wider filter width.
Therefore, both COMICS core+halo and Michelle photometries are consistent with the chopped nuclear IRS spectrum after considering difference in the filter widths.
The {\it Spitzer} and ground-based photometries match to each other despite of their different spatial resolutions, suggesting that a single compact core+halo component is located at the nucleus over the extended host-galaxy component showing the typical PAH features.
The relatively weaker PAH feature strength in the nuclear IRS spectrum (top panel of Figure~\ref{sed}; \citealt{ds10}) is apparently due to this core+halo component showing little PAH features.

\subsection{Core SED}\label{results_core}

\begin{figure}
\plotone{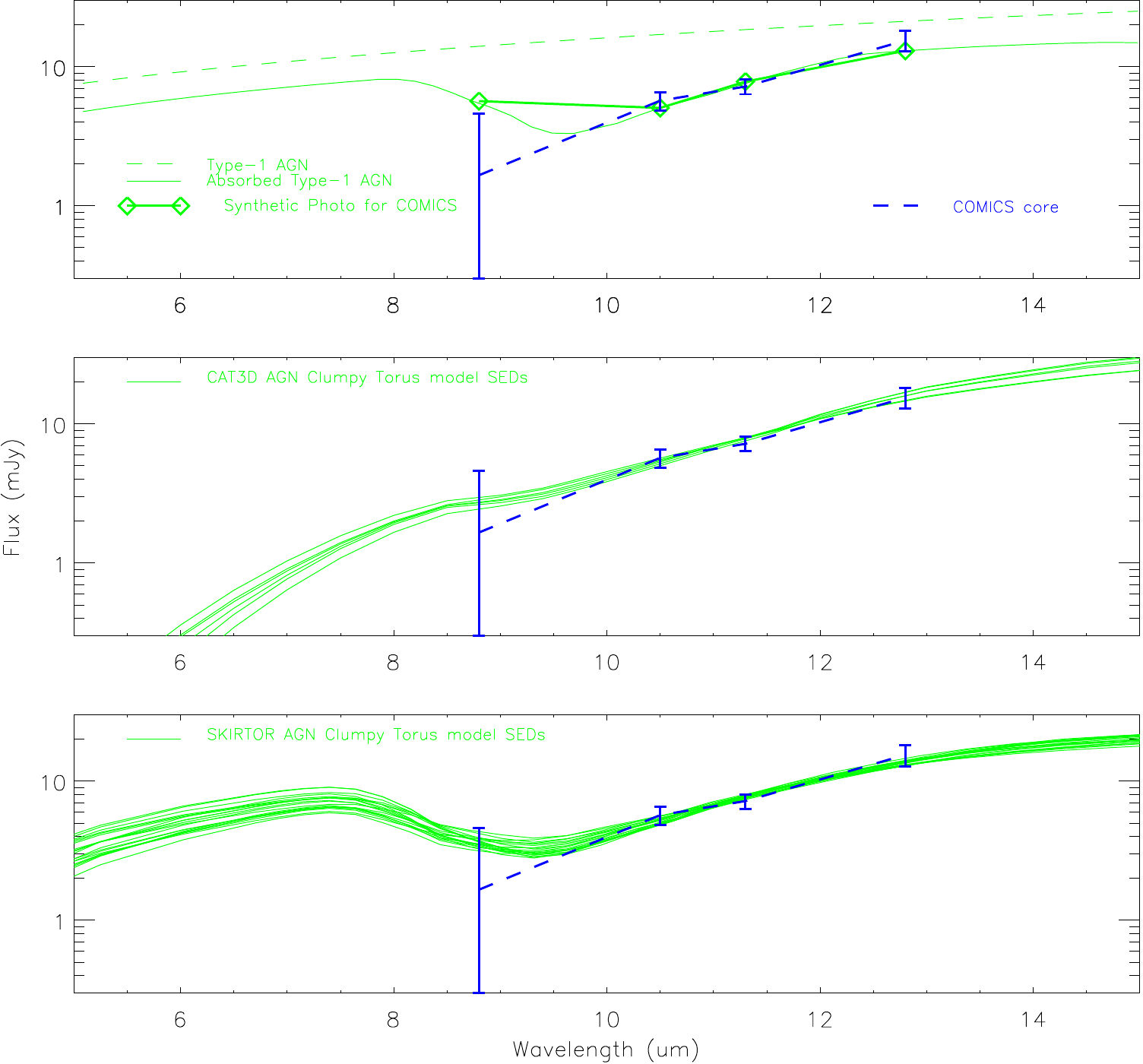}
\caption{
The same figure of Figure~\ref{sed} bottom panel but for the comparison between the COMICS core SED (blue broken line) and the AGN model SEDs of the clumpy torus.
(top)
The COMICS core SED is compared with the scaled unabsorbed type-1 AGN SED (green broken line) and an absorbed one for $\tau_{\rm 9.7}=1.5$ (green solid line).
Synthetic photometries of the absorbed AGN SED for the COMICS filters are shown with green diamonds.
(middle)
All acceptable CAT3D SED models \citep{honig10}.
(bottom)
All acceptable SKIRTOR 2.0 SED models \citep{stalevski16}.
We plot the total SED including the thermal emission from the dusty torus and the absorbed/scattered accretion-disk components, although the thermal dust emission dominates at MIR.
\label{agnsed}}
\end{figure}

The COMICS core SED can be interpreted as either the PAH-deficient chopped nuclear IRS spectrum or the AGN emission.
The bottom panel of Figure~\ref{sed} directly compares the synthetic photometries of the scaled chopped nuclear IRS spectrum with the COMICS core SED to illustrate a good match given the error bars.
Here, the scaling is to account for the flux ratio between the core and core+halo SEDs (\S \ref{radial_profile_analysis}).
This indicates that the core SED can be explained by spectrum showing little PAH features in the same way for the core+halo SED.
Alternatively, the core SED can be explained by an absorbed power-law continuum of the AGN.
To demonstrate this, we applied an absorption on type-1 AGN SED template of \cite{polletta07} by an apparent optical depth at 9.7~$\mu$m ($\tau_{9.7}$) of $\sim 1.5$ (top panel of Figure~\ref{agnsed}).
Here, we assumed simple foreground dust absorption with an extinction curve of \cite{ct06} toward the Galactic center.
We found that this model, though it is unrealistically simple, can reproduce the observation reasonably well, except for N8.8 where the photometry quality is the worst.
Note that this optical depth is much larger than that measured with IRS ($\tau_{9.7}=$0.0--0.04; \citealt{goulding12}).
This is most likely due to less contamination at the COMICS resolution by circumnuclear less-absorbed star-forming region.

We further examine the AGN models for the core SED by using the recent sophisticated AGN SED models with clumpy torus.
It has been known that the observed shallower silicate absorption at 9.7~$\mu$m even in the Compton-think AGNs can be reproduced by the AGN clumpy torus (e.g., \citealt{nenkova02}).
Recent more advanced radiative-transfer calculations for the clumpy torus medium not only successfully reproduce the MIR SEDs of both type-1 and type-2 AGNs but also provide the basis to compare with the observed spatial distribution of the infrared emission in the central parsec-scale regions of AGNs in the interferometric studies (e.g., \citealt{honig10,stalevski12,stalevski16}).
There are three groups in the model parameters: those that define the intrinsic radiation from the accretion disk, dusty torus properties such as size and density distribution as well as the dust composition, and observer's viewing angle of the torus.
The AGN luminosity sets overall flux scale of the emergent SED, not the relative SED shape, as long as the inner torus radius is set at the dust sublimation temperature, due to scaling relation of the AGN torus radiation (e.g., \citealt{nenkova08a}).
The relative SED shape is then determined mostly by overall torus shape (opening angle of the central dust-free zone of the torus and outer radius of the torus), distribution of dusts (or dust clouds) within the torus, average optical depth (or mean optical depth of each clump and the average number of the clumps) along the torus equatorial plane, as well as the viewing angle.
In particular, two most important characteristics of the observed MIR SED, the apparent depth of the silicate absorption at 9.7~$\mu$m and the overall slope (redness) of the continuum, are mostly determined by the combination of the radial dust distribution, the average optical depth along the torus equatorial plane, and the viewing angle (e.g., \citealt{honig10}).

We performed the model fitting of the core SED by using two representative AGN SED models with clumpy torus, CAT3D \citep{honig10} and SKIRTOR 2.0 (\citealt{stalevski16}; see also \citealt{stalevski12}).
These two models are very similar in overall settings of the accretion disk radiation and the torus geometry/distributions of the dusts within the torus, although there are numbers of differences in the details of the parameterization of the models, as well as the technical details of the calculations.
The AGN luminosity is estimated by using the core [Ne~{\sc ii}]-filter flux ($\nu L_{\rm \nu}$(12.3~$\mu$m)$=2.1 \times 10^{40}$ erg s$^{-1}$) and the MIR bolometric correction factor estimated by using equation (5) of \cite{gandhi09}, which is based on their MIR and X-ray luminosity correlation and the X-ray bolometric correction factor of \cite{marconi04}.
Some parameters that do not affect the MIR SED very much, such as dust composites (both in CAT3D and SKIRTOR 2.0), outer radius of the torus (in CAT3D), and fraction of the total dust mass within the clump (in SKIRTOR 2.0), are fixed to the fiducial values according to the model papers.
Middle and bottom panels of Figure~\ref{agnsed} show all acceptable model SEDs for the two models, and Table~\ref{sedfit} summarizes their parameters.
In both models, relatively larger average optical depth in the equatorial plane of the torus ($\tau_{\rm V} \times N_{\rm 0}$ in CAT3D and $\tau_{\rm 9.7}$ in SKIRTOR 2.0; see Table~\ref{sedfit}) and flatter radial distribution of the dusts (near-zero power-law index as a function of the distance from the center; $a$ in CAT3D and $p$ in SKIRTOR 2.0; see Table~\ref{sedfit}) are preferred.
This, combined with the nearly edge-on viewing angle, provides preferred conditions to see the silicate feature in absorption on the redder continuum \citep{honig10,stalevski12,stalevski16}.

\begin{table}[ht]
\caption{Fit results of the COMICS core SED with the AGN model SEDs of the clumpy torus.\label{sedfit}}
\begin{tabular}{lcc}
\tableline\tableline
parameter name and its description & parameter space & acceptable parameter range\tablenotemark{a} \\
\tableline
\multicolumn{3}{c}{CAT3D \citep{honig10}}\\
\tableline
$a$: radial dust cloud distribution index ($r^a$) & 0.0...$-2.0$ (in steps of 0.5) & 0.0..$-0.5$ (mostly 0.0) \\
$\tau_{\rm V}$: optical depth of the individual clouds at $V$ band & 30, 50, 80 & 80 \\
$N_{\rm 0}$: number of clouds along an equatorial line-of-sight & 2.5...10 (in steps of 2.5) & 7.5..10 (mostly 10.0) \\
$\theta_{\rm 0}$: half-opening angle of the torus & 5$^\circ$, 30$^\circ$, 45$^\circ$, 60$^\circ$ & 5$^\circ$..60$^\circ$ \\
$R_{\rm out}$: outer radius of the torus & 150\tablenotemark{b} & 150 \\
$i$: inclination angle\tablenotemark{c} & 0$^\circ$..90$^\circ$ (in steps of 15$^\circ$) & 60$^\circ$..90$^\circ$ (mostly 90$^\circ$) \\
dust composition & ``standard''\tablenotemark{d} ISM\tablenotemark{b} & ``standard''\tablenotemark{d} ISM \\
\tableline
\multicolumn{3}{c}{SKIRTOR 2.0 \citep{stalevski16}}\\
\tableline
$p$: radial gradient of dust density ($r^{-p}$)\tablenotemark{e} & 0.0...1.5 (in steps of 0.5) & 0.0..1.0 (mostly 0.0..0.5) \\
$q$: dust density gradient with polar angle (exp($-q~|\rm{cos}(\theta)|$))\tablenotemark{e} & 0.0...1.5 (in steps of 0.5) & 0.0..1.5 (mostly 0.0..0.5) \\
$\tau_{\rm 9.7}$: average edge-on optical depth at 9.7~$\mu$m & 3..11 (in steps of 2) & 7..11\\
$\Theta$: angle between the equatorial plan and edge of the torus\tablenotemark{f} & 10$^\circ$..80$^\circ$ (in steps of 10$^\circ$) & 50$^\circ$ \\
$R$: outer-to-inner torus radius ratio ($R_{\rm out}$/$R_{\rm in}$) & 10, 20, 30 & 20, 30 \\
$f_{\rm cl}$: fraction of total dust mass inside clumps & 0.97\tablenotemark{b} & 0.97 \\
$i$: inclination angle\tablenotemark{c} & 0$^\circ$..90$^\circ$ (in steps of 10$^\circ$) & 60$^\circ$..90$^\circ$ (mostly 80$^\circ$..90$^\circ$) \\
\tableline
\end{tabular}
\tablenotetext{a}{We show both the entire ranges of the acceptable parameters and the narrower ranges if the accepted parameters distribute mostly in the narrower ranges.}
\tablenotetext{b}{Only the SEDs with this fiducial parameter are publicly available.}
\tablenotetext{c}{$i=0^\circ$ for face-on and $90^\circ$ for edge-on viewing angles.}
\tablenotetext{d}{Standard in terms of both grain size and chemical composition. See \cite{honig10}.}
\tablenotetext{e}{$r$ and $\theta$ are coordinates in the adopted polar coordinate system. See \cite{stalevski12,stalevski16} for the details.}
\tablenotetext{f}{Half-opening angle of the dust-free cone is $90^\circ-\Theta$.}
\end{table}

\subsection{Halo SED}\label{results_halo}

The halo SED is similar to the chopped nuclear IRS spectrum, because the core and halo SEDs are similar to each other (\S \ref{radial_profile_analysis}), and the core+halo SED is consistent with the chopped nuclear IRS spectrum (\S \ref{results_core_halo}).
In particular, the [Ne~{\sc ii}] 12.8~$\mu$m emission is likely associated with the halo.
This is because the [Ne~{\sc ii}] emission is present around the nucleus at the core+halo scale as demonstrated by the difference between the COMICS core+halo SED and the Michelle SED at around 12.5--12.8~$\mu$m (\S \ref{comp_michelle}; \ref{results_core_halo}), and the core SED is most likely the dust thermal emission around the AGN as we discuss next (\S \ref{nature_core}).

\section{DISCUSSION}

\subsection{Nature of the Nuclear Unresolved Core: an Absorbed AGN or a Compact Star-Forming Cluster?}\label{nature_core}

Our upper limit of the core size and the core photometry give very high MIR surface brightness lower limit, but we cannot exclude a possibility of a very compact nuclear star cluster.
\cite{siebenmorgen08} compiled nuclear surface brightness of nearby AGNs at $\simeq 11~\mu$m measured with 8--10~m telescopes.
They demonstrated that the surface brightness of the AGNs is $> 2\times 10^4$ $L_{\rm \odot}$/pc$^2$, and showed that the surface brightness of $2\times 10^4$ $L_{\rm \odot}$/pc$^2$ is a useful diagnostic threshold to separate AGNs from compact nuclear starbursts.
The MIR luminosity of the M51 core is $\nu L_{\rm \nu}$(11.2~$\mu$m)$=1.1 \times 10^{40}$ erg s$^{-1}$ (from the core UIR11.2-filter flux), and the core size upper limit is 6~pc (FWHM; \S \ref{radial_profile_analysis}).
We found that the surface brightness is $> 1.0\times 10^4$ $L_{\rm \odot}$/pc$^2$, indicating that the core MIR surface brightness is almost comparable to that of known AGNs.
We caution that the surface brightness of the M51 core is less than the threshold by a factor of two, and very compact nuclear starbursts can have comparably large surface brightness of $\simeq 1 \times 10^4$ $L_{\rm \odot}$/pc$^2$ \citep{siebenmorgen08}.
Therefore, we cannot conclude that the AGN dominates the core in MIR emission on the basis of the MIR surface brightness argument.

We examine MIR-to-X-ray luminosity ratio of the core in an attempt to identify its nature, because both an absorbed AGN and a compact star-forming cluster with deficient PAH features can explain the MIR SED (\S \ref{results_core}).
AGNs are much more luminous in X-ray at $\gtrsim 2$~keV than star-forming clusters for given MIR luminosity (e.g., \citealt{ranalli03,ballantyne08}), and we test if the ratio of the absorption-corrected intrinsic X-ray and MIR luminosities of the core is consistent with the AGNs.
\cite{asmus15} showed that the MIR luminosity and the intrinsic X-ray (2--10~keV) luminosity corrected for the absorption based on the X-ray spectral fitting show a good linear correlation among the local AGNs (see also \citealt{horst06,horst08,horst09,gandhi09}).
They used single-dish MIR observations with 8m-class telescopes, like our observation with Subaru 8.2m telescope.
High spatial-resolution MIR photometry is essential to reduce the contamination by the circumnuclear regions and isolate the properties of the AGNs (e.g., \citealt{horst06,horst08,horst09,gandhi09}).
The correlation is found independent of the two types of the AGNs including the Compton-thick AGNs, although the Compton-thick AGNs show slightly lower MIR luminosity by $\lesssim 0.15$ dex for the same X-ray luminosity when compared to the rest.
Note that this MIR luminosity offset is smaller than the overall scatter of the correlation ($<0.4$ dex), making this correlation useful to diagnose the AGN nature of the MIR core in the Compton-thick AGN of M51.
\cite{asmus15} already included M51 in their correlation analysis by using their own Michelle photometry data \citep{asmus14}, and showed that M51 dose follow the correlation.
By using equation (2) of \cite{asmus15}, we predict the intrinsic X-ray luminosity as $L$(2--10~keV)$=1.4 \times 10^{40}$ erg s$^{-1}$ from the core 12~$\mu$m luminosity ($2.1 \times 10^{40}$ erg s$^{-1}$ from the core [Ne~{\sc ii}]-filter flux).
For comparison, the observed absorbed-corrected X-ray luminosity (converted to our assumed distance of 7.1~Mpc) is log $L$(2--10~keV)$=40.48\pm 0.6$ erg s$^{-1}$ \citep{asmus15}.
This measurement is based on multi-epoch observations with {\it Chandra} and {\it XMM-Newton} satellites, and the uncertainty includes differences among the observations (see \citealt{asmus15} and references therein).
Note that this result is consistent with the {\it BeppoSAX} measurement ($L$(2--10~keV)$=9.1 \times 10^{40}$ erg s$^{-1}$) by \cite{fukazawa01}.
We found that the observed X-ray luminosity is $0.34$ dex larger than the prediction from the MIR luminosity for general AGN populations, or is $0.19$ dex larger for the Compton-thick AGNs if we consider the above-mentioned systematic offset.
Because this difference is smaller than the uncertainty of the observed X-ray luminosity (0.6 dex), we conclude that our core MIR photometry is consistent with the MIR--X-ray luminosity correlation of the AGNs.

How can we explain discrepancy between the MIR and X-ray measurements of the hydrogen column density toward the AGN?
In the simple absorbed AGN model with foreground absorption (\S \ref{results_core}), the apparent optical depth at the silicate absorption ($\tau_{9.7} \sim 1.5$) corresponds to $N_{\rm H}\sim 3\times 10^{22}$ cm$^{-2}$.
This order-of-magnitude estimate is based on the $A_{\rm V}/\tau_{9.7}$ conversion ($=9$; \citealt{ct06}) and a standard gas-to-dust ratio within the Galactic interstellar medium ($N_{\rm H}$/$A_{\rm V} = 1.9 \times 10^{21}$ cm$^{-2}$ mag$^{-1}$).
This absorption is much smaller than expected for the Compton-thick column density measured in the X-ray ($N_{\rm H} \sim 5.6 \times 10^{24}$ cm$^{-2}$; \citealt{fukazawa01}; $N_{\rm H} \gtrsim 2 \times 10^{24}$ cm$^{-2}$; \citealt{asmus15}) in spite of considerable uncertainties in this very simple model and the conversion factors.
However, the clumpy AGN torus is known not to create very deep silicate absorption even in the Compton-thick condition (\S \ref{intro}).
In fact, we found that the model SEDs of the AGN clumpy torus (CAT3D; \citealt{honig10}; SKIRTOR 2.0; \citealt{stalevski16}) reproduce the observed core SED reasonably well (\S \ref{results_core}).
As expected, the hydrogen column density of the acceptable models is much larger than the estimate based on the apparent small optical depth at 9.7~$\mu$m:
For example, with CAT3D models, the average total optical depth in the equatorial plane of the torus is $\tau_{\rm V} \times N_{\rm 0} \simeq 800$ (Table~\ref{sedfit}), corresponding to $N_{\rm H}\sim 1.5\times 10^{24}$ cm$^{-2}$ in the edge-on viewing angle with the same standard gas-to-dust ratio.
Therefore, the AGN clumpy torus models acceptable for the MIR core SED are consistent with the X-ray observations in terms of the hydrogen column density toward the AGN.

We need to caution that, although we successfully demonstrated that the AGN clumpy torus models can reproduce the core MIR SED, there is no direct hint to suggest that the torus structure dominates the MIR emission in M51.
Even though such model SEDs are known to represent AGN SEDs in general reasonably well (e.g., \citealt{nenkova02,honig06,nenkova08a,nenkova08b,honig10,stalevski12,stalevski16}), there is mounting evidence recently to indicate that most of the MIR emission comes from the extended polar region of the AGNs, rather than from the torus as expected in the AGN unification theory (\S \ref{intro}).
In addition, although our argument based on the MIR and X-ray luminosity correlation indicates that the AGN dominates the core MIR SED, this dose not mean that the emission comes from the AGN torus.
\cite{asmus16} showed based on their 8m single-dish observations of nearby AGNs at MIR that some small fraction of their sample with sufficiently good observations shows the polar extensions, suggesting that such polar extension commonly exists in AGNs.
Because the MIR and X-ray luminosity correlation we used in our discussion is based on the total MIR luminosities of the compact nucleus including the (not well resolved) polar components \citep{asmus15}, our correlation analysis may suggest similar polar component in the M51 AGN at MIR.
Therefore, we need to wait for the future MIR interferometric observations to reveal the structure of the MIR core in M51.

\subsection{MIR Signatures of AGN Influence on Circumnuclear Region at $\lesssim 100$~pc scale}\label{agn_signature}

We found a signature of AGN influence in the nuclear IRS spectrum that shows remarkable differences from the off-nucleus one, and in the IRAC CH4/CH3 flux ratio map.
Previously, the AGN influence is found in the nuclear concentration of the hot ($T=$400--1000~K) H$_2$ gas \citep{brunner08}, the enhanced high-ionization (97.1~eV) line of [Ne~{\sc v}] (\citealt{ds10}; \S \ref{spitzer_data_analysis}), and PAH deficiency at the nucleus at the IRS resolution (\citealt{ds10}; see also \S \ref{results_core_halo}).
The PAH-deficient MIR spectrum is typically seen in AGNs (e.g., \citealt{howell07,smith07}), and is most likely due to destruction of PAH particles in the circumnuclear region due to intense hard photons from AGN and/or strong shock associated with AGN jet (e.g., \citealt{voit92,ds10}).
In addition, we found the smaller IRAC CH4/CH3 flux ratio at $\simeq 200$~pc scale around the nucleus well beyond the IRAC resolution ($\simeq 2\farcs 0$ FWHM or 70~pc; Figure~\ref{findingchart}), which is likely due to deficient PAH features in the CH4 (the 7.7 and 8.6~$\mu$m features) that are typically very bright in the star-forming regions (\S \ref{spitzer_data_analysis}).
The region with the smaller CH4/CH3 flux ratio roughly coincides with the extended radio jet (e.g., \citealt{crane92,bradley04,rampadarath15}), optical narrow-line region nebula (e.g., \citealt{cecil84,bradley04}), the soft X-ray nebula \citep{terashima01}, and the jet-entrained shocked outflowing molecular-gas clouds \citep{satoki07,satoki15}.
The IRAC flux ratio is smallest near the nucleus, where the chopped nuclear IRS spectrum shows little PAH emission (\S \ref{results_core_halo}).
These characteristics indicate more pronounced PAH destruction near the AGN.
The nuclear [Ne~{\sc v}] emission also suggests significant AGN contribution at the nucleus at the IRS resolution, because this line is usually considered as a tracer of either excitation by AGN hard photons or strong shock (e.g., \citealt{ps10}).
The observed [Ne~{\sc ii}]/[Ne~{\sc v}] flux ratio ($\simeq 4$; \S \ref{spitzer_data_analysis}) is only one tenth of the one in star-forming regions ($\sim 57$) and is much closer to the one in Seyfert galaxies ($\sim 2$; \citealt{ps10}).

What is the nature of the extended halo found in the COMICS images at a few tens~pc scale?
The halo SED is similar to the chopped nuclear IRS spectrum (\S \ref{results_halo}), suggesting the circumnuclear star formation showing little PAH emission as we discussed above.
Alternatively, the halo could be the very extended polar MIR emission of AGNs sometimes seen even with the single-dish observations (e.g., \citealt{asmus16}; \S \ref{intro}).
In fact, the MIR-to-X-ray luminosity ratio by using the combined COMICS core+halo flux, log $L$($12~\mu$m)/$L$(2--10~keV)$=0.49\pm 0.6$, is still consistent with the AGNs in general in the sample of \cite{asmus15}.
However, the former star formation model seems preferred to explain the strong [Ne~{\sc ii}] emission in the halo (\S \ref{results_halo}).
This is because strong [Ne~{\sc ii}] is typically associated with the star forming region (e.g., \citealt{ds10}), whereas the dust continuum emission dominates, without any significant [Ne~{\sc ii}] emission, in the extended polar regions in the interferometric spectroscopies of the best-studied nearby AGNs showing the polar emission (e.g., \citealt{raban09,honig12,honig13,tristram14}).
If this is the case, the halo is likely the circumnuclear star-forming regions affected by the central AGN, like in the case of the more extended region traced by the smaller IRAC CH4/CH3 flux ratio.

\subsection{Implications for a Parsec-scale Dusty Torus in M51 AGN}\label{discussion_torus}

Previous molecular line observations in mm/sub-mm wavelength have shown much smaller hydrogen column density toward the M51 nucleus than expected for the Compton-thick AGN.
This discrepancy, however, can be explained if the molecular gas is mostly associated with an one-parsec-scale torus.
\cite{satoki07} found a compact nuclear molecular-gas concentration at $0\farcs 31 \times 0\farcs 40$ resolution, which is likely associated with the MIR core, because the halo is more extended to $\simeq 1''$ FWHM.
They calculated the H$_2$ column density of $6.2 \times 10^{21}$~cm$^{-2}$ (or $N_{\rm H}\simeq 1.2 \times 10^{22}$ cm$^{-2}$) toward the nucleus.
Here, they measured the emission-line luminosity and converted it to the hydrogen number by assuming metallicity and excitation conditions of the gas.
They also assumed that the source extends just over the observing beam.
\cite{satoki15} then studied physical conditions of the molecular gas based on multi-line/multi-transition molecular-line analysis, and found that effect of gas excitation and metallicity alone cannot explain apparently very small column density reported earlier.
On the other hand, the beam dilution effect seems to significantly affect the column density measurement.
We showed in this study that the MIR core is unresolved with the 0\farcs 39 (13~pc) FWHM beam, and the one-sigma upper limit is 0\farcs 18 (6~pc) FWHM (\S \ref{radial_profile_analysis}).
If we adopt 1~pc as a fiducial torus size following statistics of nearby AGNs studied by infrared interferometry \citep{kishimoto11,burtscher13} and assume that most of the molecular gas is associated with this very compact torus, the column density is now calculated to be $\sim 1 \times 10^{24}$~cm$^{-2}$.
\cite{satoki07} already did the same calculation, but we now have better size constraint of the M51 core as well as better statistics of the AGN torus size in infrared.
\cite{satoki15} argued that mean gas density of the torus in this case is $\sim 3 \times 10^5$ cm$^{-3}$, being roughly consistent with their multi-line/multi-transition molecular-line analysis results ($\sim 10^6$ cm$^{-3}$).
Therefore, a model of one-parsec-scale, dusty, and molecular-gas-rich torus can solve the apparent column density discrepancy between the X-ray and molecular-line measurements.
Although we cannot conclude in this study that the MIR core is attributed to the AGN torus (\S \ref{nature_core}), future study of the M51 AGN at even higher spatial resolution in both infrared and sub-mm/mm wavelengths would provide us a direct view of not only the material distribution but also the fueling and feedback processes in the vicinity of the AGN.

\section{CONCLUSIONS}

We performed near-diffraction-limited ($\simeq 0\farcs 4$ FWHM resolution) $N$-band imaging of one of the nearest Compton-thick AGNs in M51 with COMICS at the Subaru telescope to study the nuclear structure and SED at 8--13~$\mu$m.
We also analyzed the archival {\it Spitzer} MIR imaging and spectroscopy data.
We then characterized properties of the AGN and the circumnuclear region at 10--100~pc scale.
Here we summarize our main findings and the implications.

We decomposed the M51 nucleus into an unresolved (at 0\farcs 39 FWHM resolution at $11~\mu$m) core and an extended halo at $\sim 1''$ scale in the $N$-band.
We estimated the intrinsic core size corrected for the instrumental effect as $< 6$~pc ($1 \sigma$).
We measured the MIR surface brightness of the M51 core at 11.2~$\mu$m to be $> 1.0\times 10^4$ $L_{\rm \odot}$/pc$^2$.
Although this surface brightness is almost comparable to that of known AGNs ($> 2\times 10^4$ $L_{\rm \odot}$/pc$^2$; \citealt{siebenmorgen08}), it is not high enough to exclude a possibility of a very compact nuclear star cluster as an origin of the core.

The core+halo SED is almost flat with slightly increasing flux at longer wavelength.
In order to interpret this SED, we made the ``chopped'' nuclear IRS spectrum to follow the COMICS chopping observation by subtracting the off-nuclear ``sky'' spectrum from the nucleus one.
This shows continuum-dominated spectrum with molecular hydrogen and [Ne~{\sc ii}] 12.8~$\mu$m emission lines but little PAH emission.
We found that the core+halo SED is reproduced by this PAH-deficient starburst spectrum.
The SED of the halo alone is similar to the core+halo SED, and is also reproduced by similar PAH-deficient spectrum with the [Ne~{\sc ii}].
In addition, we found that the IRAC 8.0~$\mu$m/5.6~$\mu$m flux ratio is smaller within $\sim 200$~pc from the nucleus, also suggesting deficient PAH 7.7 and 8.6~$\mu$m features that are typically very bright in the star-forming regions.
The PAH particles in the circumnuclear region are likely destroyed either by intense hard photons from the AGN and/or strong shock associated with the AGN jet.

The core SED is similar to the core+halo SED, and it can be reproduced by a star-forming cluster within the core showing the PAH-deficient spectrum.
Alternatively, the AGN clumpy torus model SEDs \citep{honig10,stalevski16} can reproduce the red MIR continuum with moderate apparent silicate 9.7~$\mu$m absorption ($\tau_{9.7}\sim 1.5$).
Although we cannot discriminate the two possibilities on the basis of the MIR SED examination, we found that the observed ratio of the MIR and absorption-corrected intrinsic X-ray (2--10~keV) luminosities is consistent with the one for AGNs in general.
Therefore, the core is likely dominated by the AGN in MIR emission.

\acknowledgments

Both YO and SM are supported by the National Science Council (NSC) and the Ministry of Science and Technology (MoST) of Taiwan, MOST 104-2112-M-001-034- (YO), 105-2112-M-001-024- (YO), NSC 100-2112-M-001-006-MY3 (SM), and MOST 103-2112-M-001-032-MY3 (SM).
This work is based in part on observations made with the {\it Spitzer} Space Telescope, obtained from the NASA/IPAC Infrared Science Archive, both of which are operated by the Jet Propulsion Laboratory, California Institute of Technology under a contract with the National Aeronautics and Space Administration.

{\it Facilities:} \facility{Subaru (COMICS)}, \facility{Spitzer (IRS, IRAC)}

{}

\end{document}